\documentclass{iau}
\usepackage{graphicx}
\usepackage{hyperref}
\usepackage{natbib}

\def\apj{ApJ}
\def\aj{AJ}
\def\mnras{Mon. Not. R. Astron. Soc.}

\def\apjs{Astrophys. J. Supp.}

\title[Bayesian inference of the initial conditions from large-scale structure surveys] %% give here short title %%
{Bayesian inference of the initial conditions from large-scale structure surveys}

\author[Florent Leclercq]   %% give here short author list %%
{Florent Leclercq$^{1,2,3}$}

\affiliation{$^{1}$ Institut d'Astrophysique de Paris (IAP), UMR 7095, CNRS - UPMC Universit\'e Paris 6,\\
 98bis boulevard Arago, F-75014 Paris, France
\\[\affilskip] $^{2}$ Institut Lagrange de Paris (ILP), Sorbonne Universit\'es,\\
98bis boulevard Arago, F-75014 Paris, France
\\[\affilskip]$^{3}$ \'Ecole polytechnique ParisTech,\\
Route de Saclay, F-91128 Palaiseau, France
\\[\affilskip]email: {\tt florent.leclercq@polytechnique.org}}

\pubyear{2014}
\volume{308}  %% insert here IAU Symposium No.
\pagerange{119--126}
\jname{The Zeldovich Universe: Genesis and Growth of the Cosmic Web}
\editors{R. van de Weygaert, S. Shandarin, E. Saar \& J. Einasto, eds.}
\begin{document}

\maketitle

\begin{abstract}
Analysis of three-dimensional cosmological surveys has the potential to answer outstanding questions on the initial conditions from which structure appeared, and therefore on the very high energy physics at play in the early Universe. We report on recently proposed statistical data analysis methods designed to study the primordial large-scale structure via physical inference of the initial conditions in a fully Bayesian framework, and applications to the Sloan Digital Sky Survey data release 7. We illustrate how this approach led to a detailed characterization of the dynamic cosmic web underlying the observed galaxy distribution, based on the tidal environment.
%% add here a maximum of 10 keywords, to be taken form the file <Keywords.txt>
\end{abstract}

\firstsection % if your document starts with a section,
              % remove some space above using this command.
\section{Introduction}

How did the Universe begin? This question has unusual status in physical sciences due to several profound specificities of cosmology. As the Universe is everything that exists in the physical sense, there is no exteriority nor anteriority. The experiment is unique and irreproducible, and the properties of the Universe cannot be determined statistically on a set. The energy scales at stake in the early Universe are orders of magnitude higher than anything we can reach on Earth. Finally, reasoning in cosmology is ``bottom-up" in the sense that the final state is known and the initial state has to be inferred. In the context of the cosmic web, we aim at a physical reconstruction of the pattern of initial density fluctuations that gave rise to the present network of clusters, filaments, sheets and voids. Due to the computational challenge and to the lack of detailed physical understanding of the non-Gaussian and non-linear processes that link galaxy formation to the large-scale dark matter distribution, this question has only recently been tackled. Here, we describe progress towards full reconstruction of four-dimensional state of the Universe and illustrate the use of these results for cosmic web classification in the initial and final conditions.

\section{Statistical approach: Bayesian inference}

Cosmological observations are subject to a variety of intrinsic and experimental uncertainties (incomplete observations~-- survey geometry and selection effects~--, cosmic variance, noise, biases, systematic effects), which make the inference of signals a fundamentally ill-posed problem. For this reason, no unique recovery of the initial conditions from which the present-day cosmic web originates is possible; it is more relevant to quantify a probability distribution for such signals, given the observations. Adopting this point of view for large-scale structure surveys, Bayesian probability theory offers a conceptual basis for dealing with the problem of inference in presence of uncertainty.

The introduction of a physical model in the likelihood (gravitational structure formation is the generative model for the complex final state, starting from a simple initial state -- Gaussian or nearly-Gaussian initial conditions) generally turns large-scale structure analysis into the task of inferring initial conditions \citep{JascheWandelt2013,Kitaura2013,Wang2013}. It is important to notice that this framework requires at no point the inversion of the flow of time, but solely depends on forward evaluations of the dynamical model.

Significant difficulty arises from the very large dimension of the parameter space to be explored (phenomenon usually referred to as the curse of dimensionality, \citealt{Bellman1961}). However, the problem can still be tractable thanks to powerful sampling techniques such as Hamiltonian Markov Chain Monte Carlo (HMC, \citealt{Duane1987}).

\section{Physical reconstructions}
\label{sec:borg_sdss}

The inference code \textsc{borg} (Bayesian Origin Reconstruction from Galaxies, \citealt{JascheWandelt2013}) uses HMC for four-dimensional inference of density fields in the linear and mildly non-linear regime. The physical model for gravitational dynamics included in the likelihood is second-order Lagrangian perturbation theory (2LPT), linking initial density fields (at a scale factor $a=10^{-3}$) to the presently observed large-scale structure (at $a=1$). The galaxy distribution is modeled as a Poisson sample from these evolved density fields. The algorithm self-consistently accounts for observational uncertainty such as noise, survey geometry, selection effects and luminosity dependent galaxy biases \citep{JascheWandelt2013,JascheWandelt2013b}.

In \cite{JLW2014}, we apply the \textsc{borg} code to 372,198 galaxies from the \texttt{Sample dr72} of the New York University Value Added Catalogue (NYU-VAGC, \citealt{NYUVAGC}), based ot the final data release (DR7) of the Sloan Digital Sky Survey (SDSS, \citealt{SDSS,SDSS2}).

\begin{figure}
\begin{center}
\includegraphics[width=\textwidth]{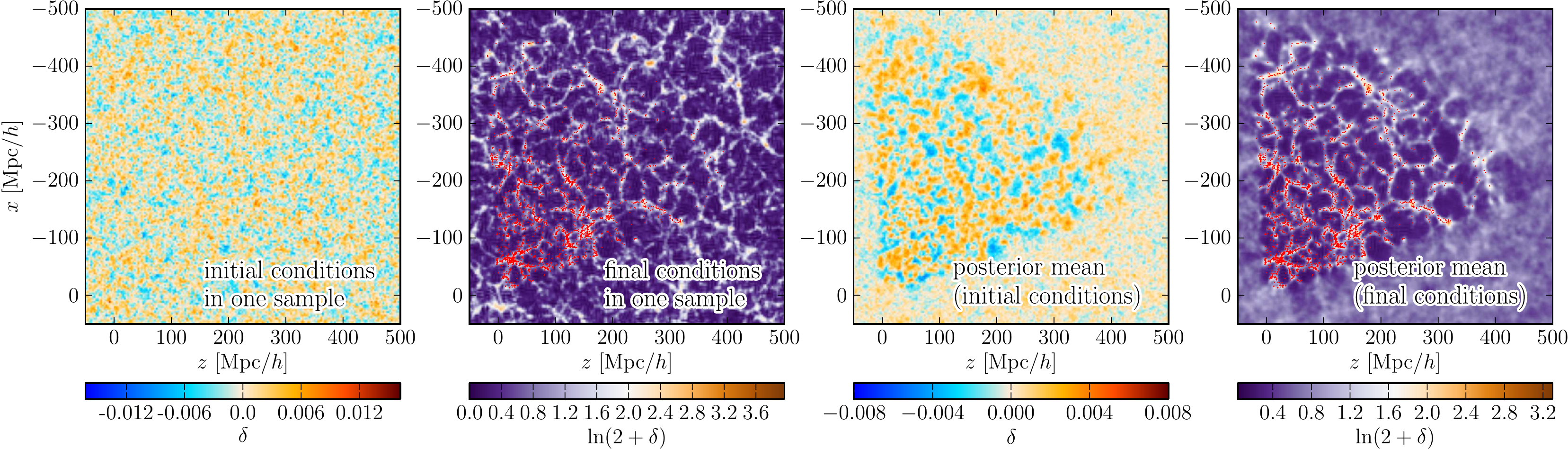}
\end{center}
\caption{Bayesian large-scale structure inference with \textsc{borg} in the SDSS DR7. Slices through one sample of the posterior for the initial and final density fields (left) and posterior mean in the initial and final conditions (right). The input galaxies are overplotted on the final conditions as red dots.}
\label{fig:borg}
\end{figure}

Each inferred sample (Fig. \ref{fig:borg}, left) is a ``possible version of the truth" for the formation history of the Sloan volume, in the form of a full physical realization of dark matter particles. The variation between samples (Fig. \ref{fig:borg}, right) quantifies joint and correlated uncertainties inherent to any cosmological observation and accounts for all non-linearities and non-Gaussianities involved in the process of structure formation. In particular, it quantifies complex information propagation, translating uncertainties from observations to inferred initial conditions.

\section{Cosmic web analysis}

\begin{figure}
\begin{center}
\includegraphics[width=\textwidth]{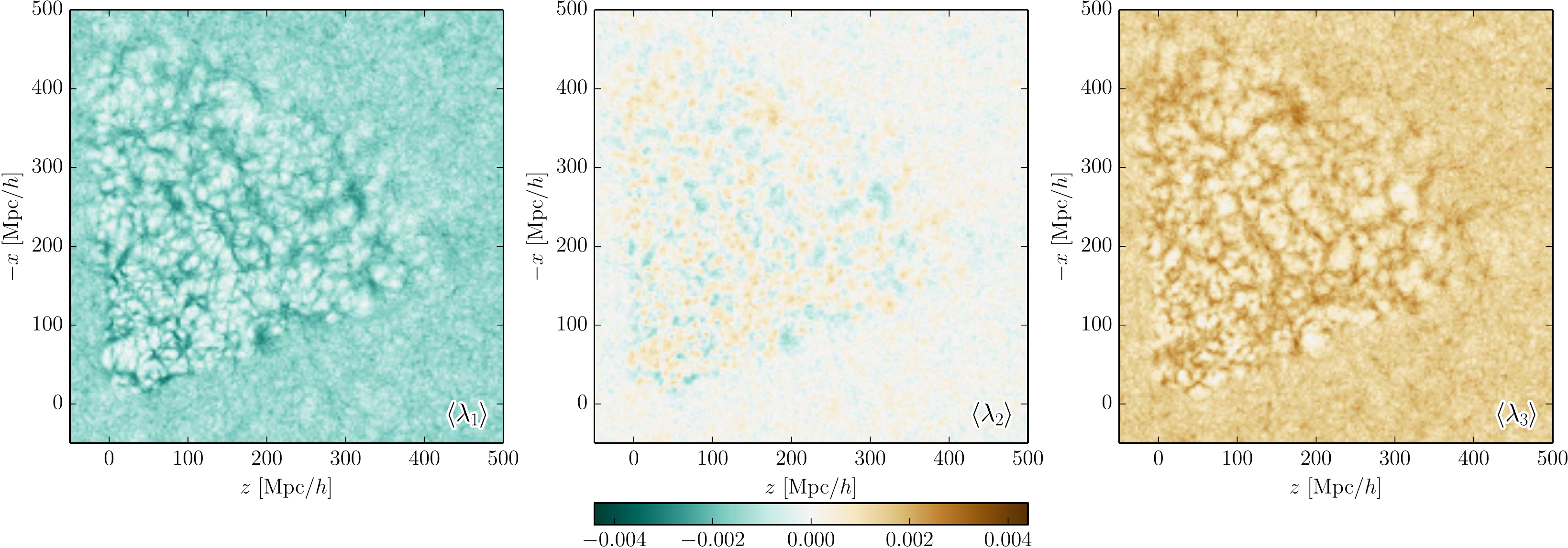}
\includegraphics[width=\textwidth]{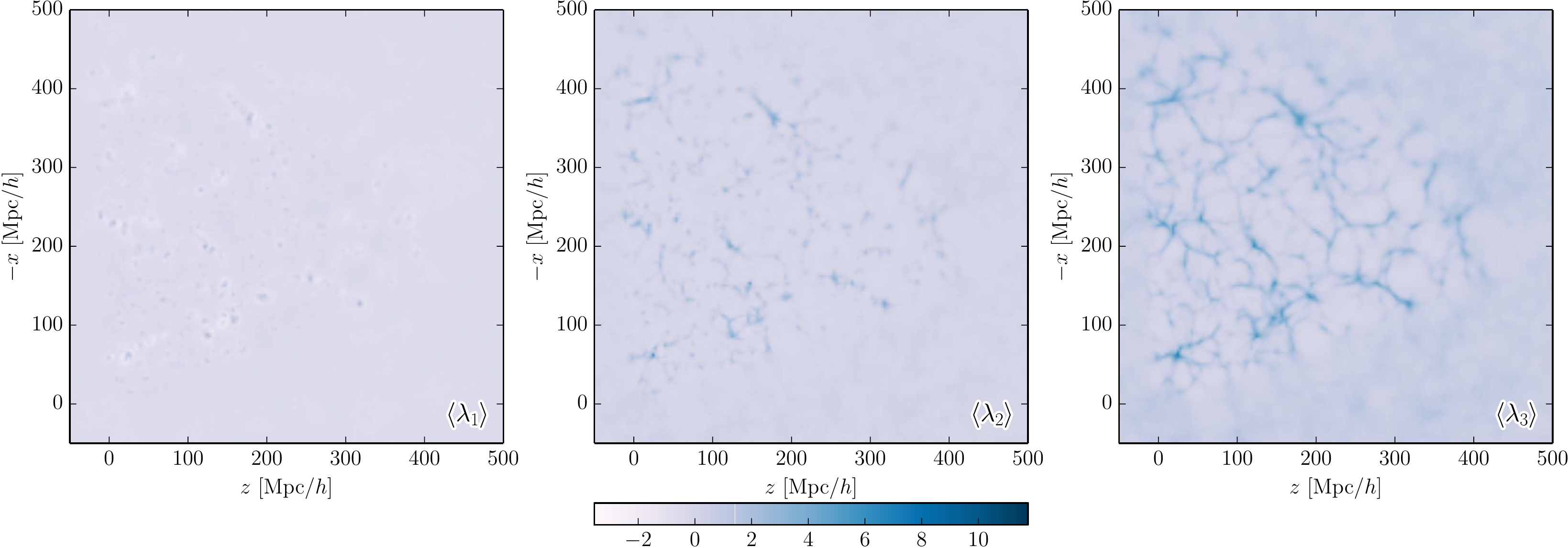}
\end{center}
\caption{Mean of the posterior pdf for the eigenvalues $\lambda_1 < \lambda_2 < \lambda_3$ of the tidal field tensor in the initial (top) and final (bottom) conditions for the large-scale structure in the Sloan volume.}
\label{fig:eigenvalues}
\end{figure}

The results presented in $\S$ \ref{sec:borg_sdss} form the basis of the analysis of \cite{LJW2014}, where we classify the cosmic large scale structure into four distinct web-types (voids, sheets, filaments and clusters) and quantify corresponding uncertainties. We follow the dynamic cosmic web classification procedure proposed by \cite{Hahn2007} (see also \citealt{Forero-Romero2009,Hoffman2012}), based on the eigenvalues $\lambda_1 < \lambda_2 < \lambda_3$ of the tidal tensor $T_{ij}$, Hessian of the rescaled gravitational potential: $T_{ij} \equiv \partial^2 \Phi / \partial \textbf{x}_i \, \partial \textbf{x}_j$, where $\Phi$ follows the Poisson equation ($\nabla^2 \Phi = \delta$). It is important to note, that the tidal tensor, and the rescaled gravitational potential are both physical quantities, and hence their calculation requires the availability of a full physical density field in contrast to a smoothed mean reconstruction of the density field. In figure \ref{fig:eigenvalues}, we show the posterior mean for $\lambda_1, \lambda_2, \lambda_3$ as inferred by \textsc{borg} is our reconstructions.

A voxel is defined to be in a cluster (resp. in a filament, in a sheet, in a void) if three (resp. two, one, zero) of the $\lambda$s are positive \citep{Hahn2007}. The basic idea of this dynamic classification approach is that the eigenvalues of the tidal tensor characterize the geometrical properties of each point in space.

\begin{figure}
\begin{center}
\includegraphics[width=\textwidth]{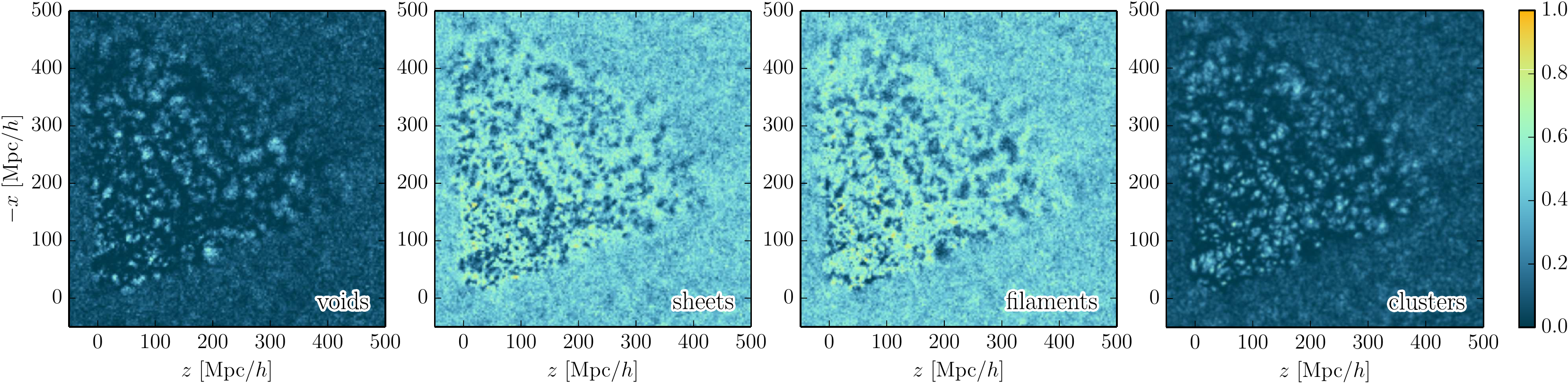}
\includegraphics[width=\textwidth]{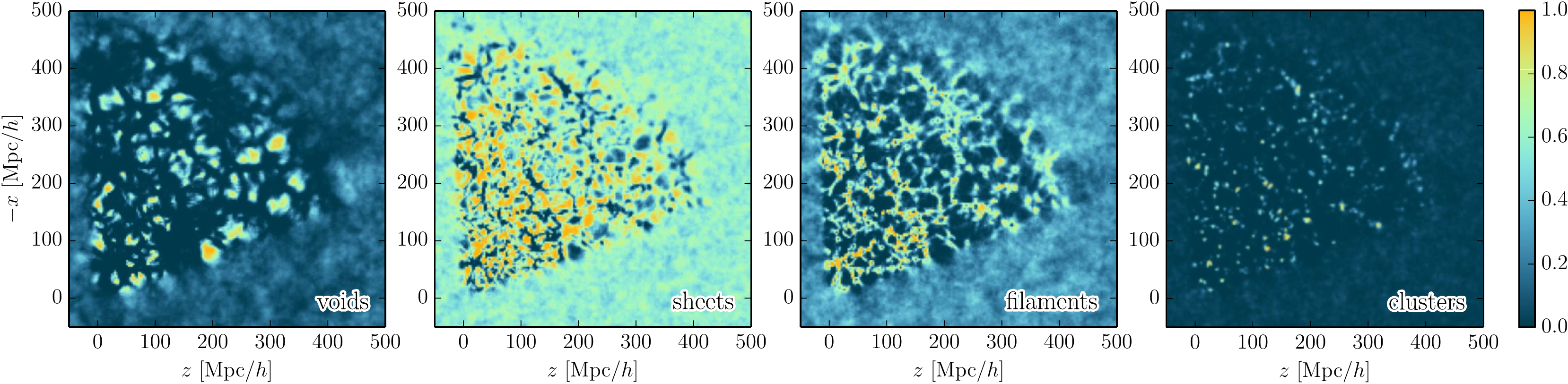}
\end{center}
\caption{Mean of the posterior pdf for the four different web-types in the initial (top) and final (bottom) conditions for the large-scale structure in the Sloan volume.}
\label{fig:ts}
\end{figure}

Our approach propagates uncertainties to structure type classification and yields a full Bayesian description in terms of a probability distribution, indicating the possibility to encounter a specific structure type at a given position in the observed volume. More precisely, by applying the above classification procedure to all density samples, we are able to estimate the posterior of the four different web-types, conditional on the observations. The mean of these pdfs are represented in Fig. \ref{fig:ts}. There, it is possible to follow the dynamic evolution of specific structures. For example, one can observe the voids expand and the clusters shrink, in comoving coordinates.

\begin{acknowledgments}
I thank Jacopo Chevallard, Jens Jasche and Benjamin Wandelt for a fruitful collaboration on the projects presented here. I acknowledge funding from an AMX grant (\'Ecole polytechnique) and Benjamin Wandelt's senior Excellence Chair by the Agence Nationale de la Recherche (ANR-10-CEXC-004-01). This work made in the ILP LABEX (ANR-10-LABX-63) was supported by French state funds managed by the ANR within the Investissements d'Avenir programme (ANR-11-IDEX-0004-02).
\end{acknowledgments}

\end{document}